\documentstyle[twoside,fleqn,espcrc2]{article}
\def\be{\begin{equation}}
\def\en{\end{equation}}
\def\bea{\begin{eqnarray}}
\def\ena{\end{eqnarray}}

\title{Magnetic Monopoles, Center Vortices and Topology of Gauge Fields
}  
\author{H. Reinhardt\thanks{talk presented by H. Reinhardt at Lattice'99, 
Pisa, Italy. }, M. Engelhardt, K. Langfeld, M. Quandt, A. Sch\"afke, 
\address{Institut f\"ur Theoretische Physik, 
Universit\"at T\"ubingen, Auf der Morgenstelle 14, D-72076 T\"ubingen, 
Germany} }

\begin{document}
\begin{abstract}
The topological properties of magnetic monopoles and center vortices arising,
respectively, in Abelian and center gauges are studied in continuum Yang-Mills
Theory. For this purpose the continuum analog of the maximum center gauge is
constructed.
\end{abstract}
\maketitle
\section{Introduction}

Recent lattice calculations have given strong evidence for two confinement
scenarios: 1. the dual Meissner effect \cite{1}, which is based on a condensate
 of
magnetic monopoles in the QCD vacuum and 2. the center vortex picture \cite{2},
 where the
vacuum consists of a condensate of magnetic flux tubes which are closed due 
to the
Bianchi identity. There are also lattice calculations which indicate that the
spontaneous breaking of chiral symmetry, which is related to the topology of
gauge fields, is caused by these objects, i.e. by either magnetic monopoles
\cite{3} or
center vortices \cite{4}. 
In this talk we would like to discuss the topological properties
of magnetic monopoles and center vortices. We will first show that in Polyakov
gauge the magnetic monopoles completely account for the non-trivial topology of
the gauge fields. Subsequently, we will extend the notion of center vortices 
to the
continuum. We will present the continuum analog of the maximum center gauge
fixing and the Pontryagin index of center vortices.

\section{Magnetic monopoles and topology}

The magnetic monopoles arise in Yang-Mills-Theories in the so called Abelian
gauges \cite{5}. For the study of these monopoles in the continuum theory the
Polyakov gauge is particularly convenient. In this gauge one diagonalizes 
the Polyakov loop  
\begin{eqnarray}
\label{1}
\Omega(\vec{x}) = Pe^{- \int^{T}_{0} dx_{0} A_{0}(x_{0}, \vec{x})} = V^{+}
\omega V
\end{eqnarray}
which fixes $V \in SU(N) / U(1)^{N-1}$ i.e. the coset part of the gauge group.
Magnetic monopoles arise as defects of the gauge fixing which occur when at 
isolated points in space $\vec{x}_{i}$ the Polyakov loop becomes 
a  center element
\begin{eqnarray}
\label{2}
\Omega(\vec{x}_{i}) = (-1)^{n_i} , \; \; \;  n_{i}\, :\, \hbox{integer} 
\end{eqnarray}
The field $A^{V} =  V A V^{+} + V \partial V^{+}$ develops then  magnetic 
monopoles. 
These monopoles have topologically quantized magnetic charge \cite{6} given by 
the winding number
\begin{eqnarray}  
\label{3}
m[V] \in \Pi_{2} (SU(2)/U(1))     
\end{eqnarray}
of the mapping $V(\vec{x})$ from a sphere $S_2$ around the magnetic 
monopole into the coset $SU(2)/U(1)$ 
of the gauge group. 

In the Polyakov
gauge the Pontryagin index can  be exactly expressed in terms of magnetic
charges. If we assume a compact space-time mani\-fold and that there are 
only point-like defects of gauge fixing,
i.e magnetic monopoles are the only magnetically charged objects arising after
gauge fixing, the Pontryagin index is given by \cite{6}
\begin{eqnarray}
\nu = \Sigma_{i} n_i m_i
\end{eqnarray}
The summation runs here over all magnetic monopoles with $m_i$ being the
magnetic charge of the monopole defined by equation (\ref{3}) and the integer 
$n_i$ is defined by the value of the Polyakov-loop at the monopole position
(\ref{2}).
This relation shows that the magnetic monopoles completely account for the 
non-trivial topology of gauge fields, at least in the Polyakov gauge. 
Unfortunately,
in other Abelian gauges like maximum Abelian gauge, such a simple relation
between Pontryagin index and topological charges is not yet known. However, in
the maximum Abelian gauge correlations between instantons and monopoles has been
found, in both analytical and lattice studies \cite{3}. 

\section{Center vortices in the continuum}

In order to
extend the notion of center vortices to the continuum theory let us, 
undertake a detour through the lattice by putting a given smooth gauge field 
$A_{\mu}(x)$
on a fine lattice in the standard fashion by introducing the link variables
$U_{\mu}(x) = exp(-aA_{\mu}(x))$. For initially smooth gauge fields $A_{\mu}(x)$
and for sufficiently fine lattices all link variables will be close to unity,
which represents the ''north pole'' of the SU(2) group manifold $S_3$. 
However, in the process of gauge fixing,  gauge transformations are performed
which transform some of the links from the northern to the southern hemisphere. In fact, we can separate each gauge transformation into a transformation 
which merely switches a link from the northern to the southern
hemisphere (or vice versa) and one which rotates link variables inside 
either the northern or the southern hemisphere but does not switch hemispheres.

In the maximum center gauge condition
\be \sum_{x,{\mu}} (T_{r}U_{\mu}(x))^{2} 
\to max \; , 
\en 
which is obviously insensitive to center gauge transformations, one exploits
gauge transformations to 
rotate a link variable as close as possible to a center element. 
Once the maximum center gauge has been implemented, center
projection implies to replace all links by their closest center element. 
One obtains then a
$Z(2)$ lattice which contains $D - 1$ dimensional hypersurfaces $\Sigma$  on 
which all links take a non-trivial center element that is $U = -1$ in the 
case of
$SU(2)$. The $D - 2$ dimensional boundaries $\partial\Sigma$ of the 
hypersurfaces
$\Sigma$ represent the center vortices, which, when non-trivially linked to a
Wilson loop, yield a center element for the latter.

A carefully analysis shows that the continuum analogies of the center 
vortices are defined  by the gauge potential \cite{7},
\begin{eqnarray}
\label{4}
{\cal A}_{\mu}(x,\Sigma) = E \int_{\Sigma} d^{D-1} \tilde{\sigma}_{\mu} \delta^{D}(x -
\bar{x}(\sigma))      
\end{eqnarray}
where $d^{D-1} \tilde{\sigma}_{\mu}$ is the dual of the $D-1$
dimensional volume element. Furthermore, the quantity $E = E_{a}H_{a}$ with
$H_{a}$ being the generators of the Cartan algebra represents 
(up to a factor of
$2\pi$) the so called co-weights which satisfy $exp (-E) = Z \in Z(N)$.
Due to this fact the Wilson-loop calculated from the gauge potential (\ref{4})
becomes,
\begin{eqnarray}
\label{5}
W[{\cal A}](C) = \exp (- \oint_{C} {\cal A}) = Z^{I(C,\Sigma)}
\end{eqnarray}
where $I(C,\Sigma)$  is the intersection number between the Wilson-loop $C$ and
the hypersurface $\Sigma$. The representation, (\ref{4}), 
is referred to as ideal
center vortex. One should emphasize that the hypersurface $\Sigma$ can be
arbitrarily deformed by a center gauge transformation keeping, however, 
its boundary
$\partial\Sigma$, i.e. the position of the center vortex, fixed. Thus for fixed 
$\partial\Sigma$  the dependence of the gauge potential (\ref{4}) on the 
hypersurface itself is a gauge artifact. 

The dependence on the hypersurface $\Sigma$ can be removed by performing the
gauge transformation 
\begin{eqnarray}
\label{6}
\varphi(x,\Sigma) = exp (- E \Omega (x,\Sigma))
\end{eqnarray}
where $\Omega(x,\Sigma)$ is the solid angle subtended by the hypersurface
$\Sigma$ as seen from the point $x$. One finds then
\begin{eqnarray}
\label{7}
{\cal A}_{\mu}(x,\partial\Sigma) = \varphi (x,\Sigma) \partial_{\mu}
\varphi^{+}(x,\Sigma) + a_{\mu} (x,\partial\Sigma)
\end{eqnarray}
where
\begin{eqnarray}
\label{8}
a_{\mu}(x, \partial\Sigma) = E \int_{\partial\Sigma} d^{D-2}
\tilde{\sigma}_{\mu\nu} \partial_{\nu} D (x - \bar{x}(\sigma))    
\end{eqnarray}
depends only on the vortex position $\partial\Sigma$ and is referred to 
as ''thin
vortex''. Here $D(x -\bar{x}(\sigma))$ represents the Green function 
of the $D$
dimensional Laplacian. In fact, one can show \cite{7} that the thin vortex 
represents the
transversal part of the ideal vortex $a_{\mu}(x, \partial\Sigma) = P_{\mu\nu}
{\cal A}_{\nu} (x, \Sigma)$ where $P_{\mu\nu} = \delta_{\mu\nu} - \frac
{\partial_{\mu}\partial_{\nu}}{\partial^{2}}$ is the usual transversal
projector.
A careful and lengthy analysis \cite{7} yields then the conclusion that 
the continuum
analog of the maximum center gauge fixing is defined by 
\begin{eqnarray}
\label{9}
\min\limits_{\partial\Sigma} \min\limits_{g} \int \hbox{Tr}  (A^{g} - 
a (\partial\Sigma))^{2}
\end{eqnarray}
where the minimization is performed with respect to all (continuum) gauge
transformations $g \in SU(2)/Z(2)$ (which represents per se coset gauge
transformations) and with respect to all vortex surfaces $\partial\Sigma$. For
fixed thin center vortex field configuration $a (\partial\Sigma)$ 
minimization with
respect to the continuum gauge transformation g yields the background gauge
condition 
\begin{eqnarray}
\label{10}
\left[\partial_{\mu} + a_{\mu} (\partial\Sigma), A_{\mu} \right]
 = 0
\end{eqnarray}
where the thin vortex field $a_{\mu} (x, \partial\Sigma)$ figures as background
gauge field. One should emphasize, however, that the background field has to be
dynamically determined for each given gauge field $A_{\mu} (x)$ and thus depends
on the latter.

Obviously in the absence of a vortex structure in a considered gauge field 
$A_{\mu} (x)$ the background gauge condition reduces to the Lorentz gauge
$\partial_{\mu}  A_{\mu} = 0$.

\section{Topology of Center vortices}

Once  the center vortex configurations 
in the continuum are at our disposal, it 
is straightforward to calculate their Pontryagin index. 
In the continuum formulation where center vortices live in the
Abelian subgroup by construction the direction of the magnetic flux of 
the vortices is fully kept.
The explicit calculation \cite{7} 
shows that the Pontryagin index $\nu$ of the center vortices
is given by $\nu = \frac{1}{4}
I(\partial\Sigma,\partial\Sigma)$
where $I(\partial\Sigma,\partial\Sigma)$ represents the self-intersection number of the closed vortex sheet
$\partial\Sigma$. A careful analysis shows that for closed orientable surfaces
the self intersection number vanishes. In order to have a non-trivial Pontryagin
index the vortex surfaces have to be not globally orientable, i.e., they have
to consist of orientable pieces. One can further show that at the border between
orientable vortex patches magnetic monopole currents flow. It is these monopole
currents which make the vortex sheet non orientable since they change the
orientation of the magnetic flux. Thus we obtain that even for the center
vortices the non-trivial topology is generated by magnetic monopole currents
flowing on the vortex sheets. This is consistent with our finding in the
Polyakov gauge where the Pontryagin index was exclusively expressed in terms of
magnetic monopoles \cite{6}.

By implementing the maximum center gauge condition in the continuum one 
can derive, in an approximate
fashion, an effective vortex theory, where the vortex action can be 
calculated in
a gradient expansion. The leading order gives the Nambu-Goto action while in
higher orders curvature terms appear. A model based on such an effective vortex
action, in fact, reproduces the gross features of the center vortex picture 
found in numerical lattice simulations.

\section*{ Acknowledgment:}

This work was supported in part by the DFG-Re 856/4-1 and DFG-En 415/1-1.

	
\end{document}